# Switching Magnetism and Superconductivity with Spin-Polarized Current in Iron-Based Superconductor

Seokhwan Choi,[1] Hyoung Joon Choi,[2] Jong Mok Ok,[3,4] Yeonghoon Lee,[1] Won-Jun Jang,[1,5,ǂ] Alex Taekyung Lee,[6] Young Kuk,[7] SungBin Lee,[1] Andreas J. Heinrich,[8,9] Sang-Wook Cheong,[10] Yunkyu Bang,[11] Steven Johnston,[12] Jun Sung Kim,[3,4] and Jhinhwan Lee[1]*

[1] *Department of Physics, Korea Advanced Institute of Science and Technology, Daejeon 34141, Korea*

[2] *Department of Physics and Center for Computational Studies of Advanced Electronic Material Properties, Yonsei University, Seoul 03722, Korea*

[3] *Department of Physics, Pohang University of Science and Technology, Pohang 37673, Korea*

[4] *Center for Artificial Low Dimensional Electronic Systems, Institute for Basic Science, Pohang 37673, Korea*

[5] *Center for Axion and Precision Physics Research, Institute for Basic Science (IBS), Daejeon 34051, Korea*

[6] *Department of Applied Physics and Applied Mathematics, Columbia University, New York 10027, USA*

[7] *Department of Physics and Astronomy, Seoul National University, Seoul 151-747, Korea*

[8] *Center for Quantum Nanoscience, Institute for Basic Science (IBS), Seoul 03760, Korea*

[9] *Physics Department, Ewha Womans University, Seoul 03760, Korea*

[10] *Rutgers Center for Emergent Materials and Department of Physics and Astronomy, Rutgers University, Piscataway, New Jersey 08854, USA*

[11] *Department of Physics, Chonnam National University, Kwangju 500-757, Korea*

[12] *Department of Physics and Astronomy, University of Tennessee, Knoxville, Tennessee 37996-1200, USA*

ǂ *Present address: Center for Quantum Nanoscience, Institute for Basic Science (IBS), Seoul 03760, Republic of Korea; Department of Physics, Ewha Womans University, Seoul 03760, Republic of Korea*

Correspondence*: jhinhwan@kaist.ac.kr .

**Abstract**

We have explored a new mechanism for switching magnetism and superconductivity in a magnetically frustrated iron-based superconductor using spin-polarized scanning tunneling microscopy (SPSTM). Our SPSTM study on single crystal $Sr_2VO_3FeAs$ shows that a spin-polarized tunneling current can switch the Fe-layer magnetism into a non-trivial $C_4$ (2×2) order, which cannot be achieved by thermal excitation with unpolarized current. Our tunneling spectroscopy study shows that the induced $C_4$ (2×2) order has characteristics of plaquette antiferromagnetic order in the Fe layer and strongly suppresses superconductivity. Also, thermal agitation beyond the bulk Fe spin ordering temperature erases the $C_4$ state. These results suggest a new possibility of switching local superconductivity by changing the symmetry of magnetic order with spin-polarized and unpolarized tunneling currents in iron-based superconductors.

**Main text**

Iron-based superconductors (FeSCs) have shown intriguing phenomena related to the coexistence of magnetism and superconductivity below the superconducting transition temperature ($T_c$) [1-3]. Although understanding of their detailed interplay is still under debate, certain magnetic orders seem to be very crucial in realizing coexistent superconductivity [3-15]. Recent studies have shown new re-entrant $C_4$ symmetric antiferromagnetic phases ($C_4$ magnetism from now on) coexisting with superconductivity, and have reported that the superconducting $T_c$ is suppressed by $C_4$ magnetic order [16-19]. Direct atomic scale control of the Fe layer's magnetic symmetry and the determination of its correlation with superconductivity may be useful for an in-depth understanding of the interplay between superconductivity and magnetism. To our knowledge, there has been no report of direct real-space observation of such a control by local probes and atomic scale demonstration of the correlation of magnetism and superconductivity.

In this regard, the parent compound tetragonal iron-based superconductor $Sr_2VO_3FeAs$ with $T_c \approx 33$ K [20] is an ideal candidate where the interplay between magnetism and superconductivity can be directly demonstrated due to its nearly degenerate magnetic ground states. $Sr_2VO_3FeAs$ has two types of square magnetic ion lattices: a square Fe lattice in the FeAs layer and a square V lattice in the two neighboring $VO_2$ layers. At optimal doping, the FeAs layer usually prefers $C_2$ magnetism harboring superconductivity while the $VO_2$ layer prefers $C_4$ magnetism [1-3,21]. Previous experimental studies of $Sr_2VO_3FeAs$ [22-28], however, have reported inconsistent results about magnetic order; recent nuclear magnetic resonance (NMR) measurements on single crystals [29] and neutron diffraction [30] experiments show that there is no long range magnetic order in the V lattice at any temperature while in the Fe lattice a magnetic order with a small moment of ~ 0.05 $\mu_B$, possibly due to frustration, is developed below 50 K. Indeed, a theoretical GGA calculation has suggested that there can be a number of competing metastable magnetic states composed of different symmetries in V and Fe layers [21]. This is a reasonable theoretical prediction considering the coupling and frustration between V and Fe layers (Supplemental Material Sec. II). Therefore, it has been quite a challenging and interesting experimental task to determine the possible magnetic ground states of the heterostructure superconductor $Sr_2VO_3FeAs$ and the possible methods to adjust their balances.

One possible way to explore the potentially frustrated magnetic states and their relation to superconductivity is using a spin-polarized scanning tunneling microscope (SPSTM) to locally modify the magnetic environment with spin-polarized tunneling current. Our density functional theory (GGA) calculation (Supplemental Material Sec. I) showed a possibility that a non-zero net spin density by the injection of spin-polarized tunneling current can induce a $C_4$ magnetic order from a pristine $C_2$ magnetic order due to the Hund's interaction, as illustrated schematically in Fig. 1(a)-(d). The spin transfer torque and Joule heating effects will then provide the energies to overcome the characteristic potential barriers between the different magnetic states [31].

In this letter, using a SPSTM we demonstrate that a spin-polarized tunneling current can induce a nontrivial metastable $C_4$ magnetic order in the Fe layer not usually achievable through thermal excitation. We also showed that a thermal annealing beyond the bulk Fe magnetic ordering temperature erases the induced $C_4$ magnetic order. From the tunneling spectroscopy analysis measured inside and outside of the region of the induced $C_4$ magnetic order, we also find a signature of suppressed superconductivity in the $C_4$ order region, which is shown to be consistent with the nesting and spin fluctuation scenario of iron-based superconductivity.

We grew single crystals of $Sr_2VO_3FeAs$ with a self-flux method [29], which are then cleaved *in situ* at temperature ~15 K just before mounting on the STM head. Due to the weakly Van der Waals-coupled SrO-SrO layers, the cleaved surface is almost always terminated with symmetrically cleaved SrO layer. For real-space magnetic imaging and injection of spin-polarized current, we have developed a technique of SPSTM with antiferromagnetic Cr-cluster tip (Cr tip, from now on), which is created *in situ* on a Cr(001) surface (Supplemental Material Sec. III). Each Cr tip is confirmed on Cr(001) steps for spin contrasts [Fig. S2(c)] and no gap in the *dI/dV* spectrum on Cr(001) [Fig. S2(e)].

The 4.6 K STM topographic image of the as-cleaved SrO top layer of $Sr_2VO_3FeAs$ taken with an unpolarized W tip in Fig. 2(c) shows small randomly oriented domains of quasi-$C_2$-symmetric atomic corrugations. These show no preference for any particular four-fold lattice direction over large scales, consistent with their identity as surface reconstructions (SR) in the absence of bulk orthorhombicity [30].

In contrast, our SPSTM images with spin-polarized Cr tip show (above a small bias threshold [~30 meV, ~25 pA]), a $C_4$ symmetric (2×2) order with intra-unit-cell topographic modulations [Fig. 2(d)-(e)] without any signature of SR seen in unpolarized tip images [Fig. 2(c)]. This observation implies that the spin-polarized current induces randomly fluctuating SR with a flat time average (See Fig. S7). At the same time, any magnetic signal of Fe-layer observed on the top layer oxygen should be the average of the four neighboring Fe spins connected to the As ions in each vertical O-V-As tunneling path as shown in Fig. 2(a) and (b). Hence the most natural explanation for the observed (2×2) pattern with three groups of apparent height levels is the plaquette order in the Fe lattice with flat time-averaged SR. The Fourier-transformed $q$-space image [the inset of Fig. 2(d)] also shows the double wave vectors $\boldsymbol{Q} = \left(\frac{\pi}{2}, \frac{\pi}{2}\right)_{Fe}$ and $\boldsymbol{Q}^* = \left(\frac{\pi}{2}, -\frac{\pi}{2}\right)_{Fe}$ expected from the plaquette order in Ref. 8.

To understand the nature of magnetic metastability in this system, we performed a comparative study of bias-dependent topographic measurements using unpolarized (W) and spin-polarized (Cr) tips at 4.6 K. Using the unpolarized tip, shown in Fig. 3(a)-(c), we found that the surface starts to change at biases beyond $V_{th}^{N} \approx$ 300 meV and the fluctuation becomes so rapid above 400 meV that the surface starts to appear essentially flat as a result of time-averaging of the fluctuations. Returning to the low bias condition, as shown in Fig. 3(d), we observe that the square area which experienced the high bias scanning has completely changed with sharply defined boundaries, as shown in Fig. 4(a).

In case of the spin-polarized tip [Fig. 3(e)-(g)], the SPSTM image is qualitatively identical to the unpolarized tip case at low bias near 10 meV, but the surface starts to change beyond a threshold bias $V_{th}^{SP} \approx$ 30 meV, revealing the (2×2) domain structure and its phase domain walls. The significantly lower bias threshold voltage for a spin-polarized tip is indicative of a final state qualitatively different from that achieved by an unpolarized tip. Returning to the very low bias condition shown in Fig. 3(h), we found that the $C_4$ order is still retained with extra fluctuations (visible as random horizontal streaks) implying the extra degeneracy in the $C_4$ state.

The qualitative equivalence of the pristine surface images taken with an unpolarized tip [Fig. 3(a)] and a spin-polarized tip [Fig. 3(e)] can be understood from the fact that the pristine state will probably have either $C_2$ single stripe correlations or (in the presence of disorder) short-range $C_2$ single stripe orders, both supporting superconducting pairing (Refs. 21, 29, and Supplemental Material Sec. I). None of these two kinds of $C_2$ magnetism is detectable by SPSTM due to the particular tunneling geometry of this material [Fig. S3(d)].

In order to explore the possibility of erasing of the $C_4$ order by thermal excitation, we performed a variable temperature Cr-tip SPSTM measurement [Fig. 3(i)-(l) and Fig. S10]. We found that the $C_4$ order can be erased near 60 K, right above the Fe magnetic ordering temperature found in NMR measurement [29]. On the other hand, application of magnetic field up to 7 T does not induce any qualitative change in the $C_4$ (2x2) pattern in the Cr tip SPSTM topograph [Fig. S11]. These show that the induced $C_4$ order is an antiferromagnetic order in the Fe layer and the switching of the Fe magnetism is reversible by thermal agitation beyond the bulk Fe magnetic ordering temperature.

To study the connection between superconductivity and the $C_4$ magnetic order, we performed a comparative spectroscopic study. We first acquired a large area topograph using a unpolarized tip with bias condition below threshold $V_{th}^{N}$. We then scanned over a smaller square area near the center [black dotted square in Fig. 4(a)] with bias condition exceeding the threshold $V_{th}^{N}$, simulating thermal annealing in this area. Figure 4(a) shows the topograph taken immediately afterwards with a bias condition below $V_{th}^{N}$. It shows the changed surface topographic pattern, which corresponds to another instance of the nearly degenerate ground states achievable by tunneling current-induced non-uniform thermal excitation. Then we measured the *dI/dV* spectra inside [annealed, Fig. 4(c), blue solid curve] and outside [as-cleaved, Fig. 4(c), green solid curve] of the central high-bias-scanned region. The tunneling spectra measured in both regions identically show various features; a pair of superconducting coherence peaks near $-6$ meV and +6 meV, and a SDW-gap-edge-like features near $-18$ meV and +14 meV. These spectral features are virtually independent of the changes in SRs as demonstrated in Supplemental Material of Ref. 32. This implies that most of the spectral features, including the superconducting gap, are the physics in the FeAs layer beneath the topmost $Sr_2VO_3$ layer [32].

In the case of spin-polarized (Cr) tip, the results are qualitatively different. Fig. 4(b) shows a large area topograph taken with bias condition below $V_{th}^{SP}$ after scanning over the smaller square region (black dotted square) with biases over $V_{th}^{SP}$ [Fig. 3(e)-(g)]. The central square region shows the well-defined $C_4$ domains (and various domain walls) induced by the spin-polarized current. The *dI/dV* spectra measured in the region with $C_4$ order [red and purple curves in Fig. 4(d)] shows that the superconducting coherence peaks and the SDW-gap-edge-like features are both significantly suppressed in the presence of $C_4$ magnetic order.

One plausible explanation for suppressed superconductivity in this particular $C_4$ (plaquette) order is related to the mutual relationship of the spin wave dispersion in Fig. 2(f) (derived from Ref. 8) and the overlaid Fermi-surfaces observed in angle-resolved photoemission spectroscopy (ARPES) measurement [33]. For the $C_4$ plaquette order, the low energy spin fluctuations with wave vectors $\boldsymbol{Q}$ and $\boldsymbol{Q}^*$ do not satisfy the nesting condition between any pair of the Fermi surfaces Γ and M (X) and thus are unable effectively mediate pairing in the spin fluctuation-based theory of iron-based superconductivity. According to this scenario, the suppression of the nesting condition by the induced $C_4$ plaquette order will have a more drastic effect on superconductivity compared with switching between $C_2$ and $C_4$ orders that maintain the nesting conditions, as shown in the recent studies on $Ba_{1-x}K_xFe_2As_2$ [17] and $Ba_{1-x}Na_xFe_2As_2$ [18] where a more subtle $T_c$ reduction was observed. Among multiple theories of iron-based superconductivity based on spin fluctuations [34,35] and orbital fluctuations [36,37], our experimental results on this material seem to favor the former.

In summary, we carried out a real-space study of correlation between superconductivity and $C_4$-magnetism in an iron-based superconductor by changing the magnetic symmetry using spin-polarized STM. In this magnetically frustrated material, a spin-polarized tunneling current induced a nontrivial metastable $C_4$ order not usually accessible through thermal excitation, while thermal agitation beyond the bulk Fe spin ordering temperature erased the $C_4$ state. We also observed suppressed superconductivity in the $C_4$ order region induced by spin-polarized current consistent with the spin fluctuation-based theories. These are a unique and clear demonstration of switching the Fe-layer magnetism and superconductivity by spin-polarized current injection and thermal agitation. As suggested in Fig. S12, our findings may be extended toward future studies for heterostructure superconductor devices manipulating magnetism and superconductivity using spin-polarized and unpolarized currents.

## Acknowledgement

The authors are thankful for helpful discussions with A. Chubukov, I. Mazin, H.-J. Lee, S.-J. Kahng, Ja-Yong Koo, C. Kim, J.J. Yu, W. Wu, K.-J. Kim, and J.H. Shim. This work was supported by National Research Foundation (NRF) through the Pioneer Research Center Program (No. 2013M3C1A3064455), the Basic Science Research Programs (No. 2017R1D1A1B01016186), SRC Center for Topological Matter (No. 2011-0030785), the Creative Research Initiative Program (No. 2011-0018306), the Max Planck POSTECH/KOREA Research Initiative Program (No. 2011-0031558), the Brain Korea 21 PLUS Project of Korea Government. It is also supported by IBS-R017-D1, IBS-R027-D1 and by Korea Research Institute of Standards and Science through the Metrology Research Center Program funded (No. 2015-15011069), by MSIP of Korea through NRF (2015R1C1A1A01052411), by the Samsung Advanced Institute of Technology (SAIT), and by the Gordon and Betty Moore Foundation's EPiQS Initiative through Grant GBMF4413 to the Rutgers Center for Emergent Materials. Computational resources have been provided by KISTI Supercomputing Center (Project No. KSC-2016-C3-0052).

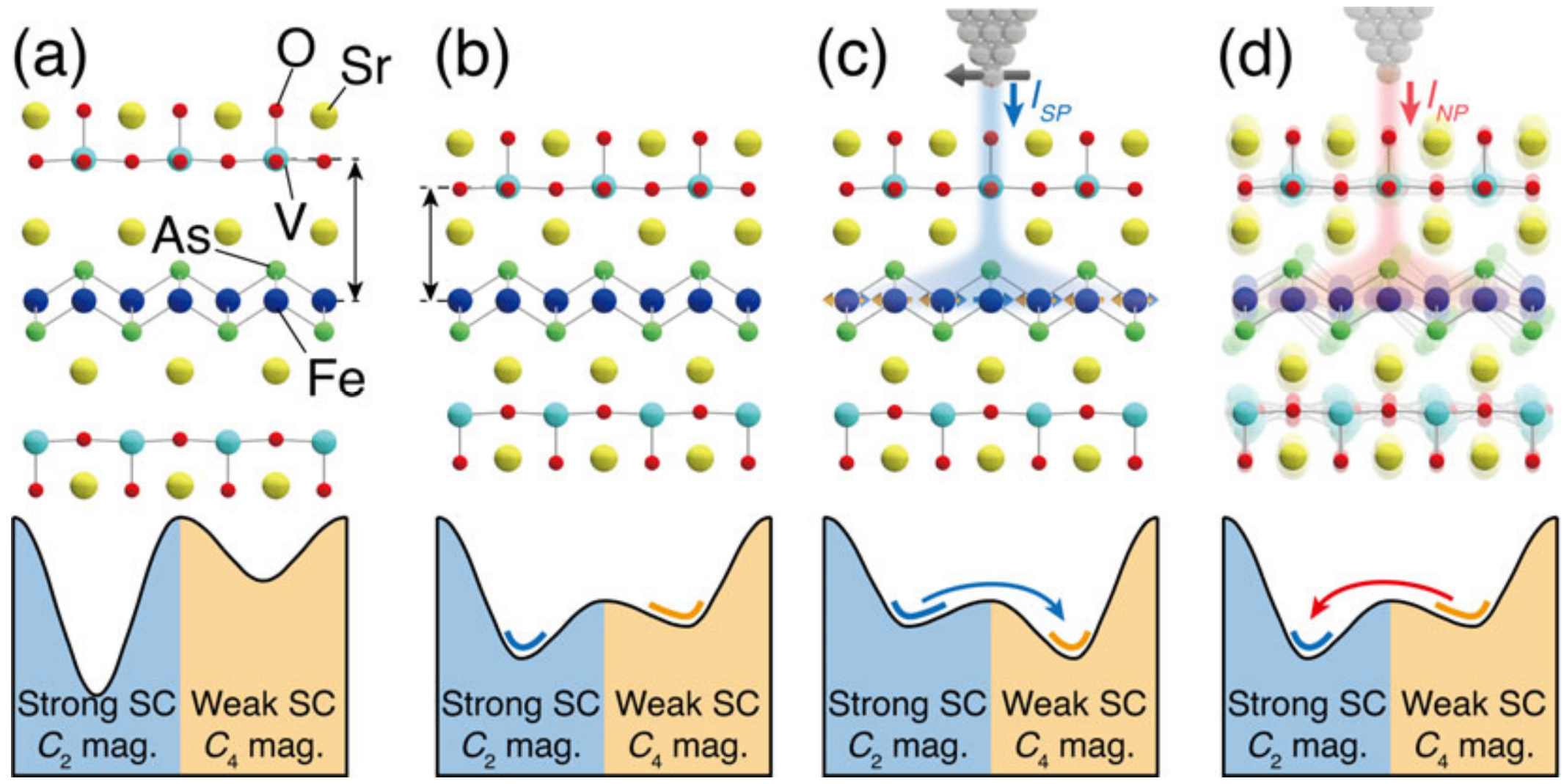

FIG. 1. (a)-(d) Schematic illustrations of FeAs-layer configuration potential landscapes for $Sr_2VO_3FeAs$ in various situations: (a) The imaginary case of FeAs and $Sr_2VO_3$ layers being separated sufficiently apart while the electron doping from Sr layer retained near optimal. The $C_2$ magnetism in Fe layer with strong superconductivity is preferred. (b) The natural separation found in $Sr_2VO_3FeAs$ results in inter-layer coupling and near degeneracy among the magnetic states with different symmetries, with the $C_2$ magnetism with strong superconductivity still being the ground state. (c) If a sufficiently strong spin-polarized current is injected, the balances among these states may change, possibly resulting in $C_4$ magnetic states with weak superconductivity in the FeAs layer. (d) When the sample is thermally annealed globally or locally with high bias tunneling current injection, it may return to the ground states with $C_2$ magnetism and strong superconductivity.

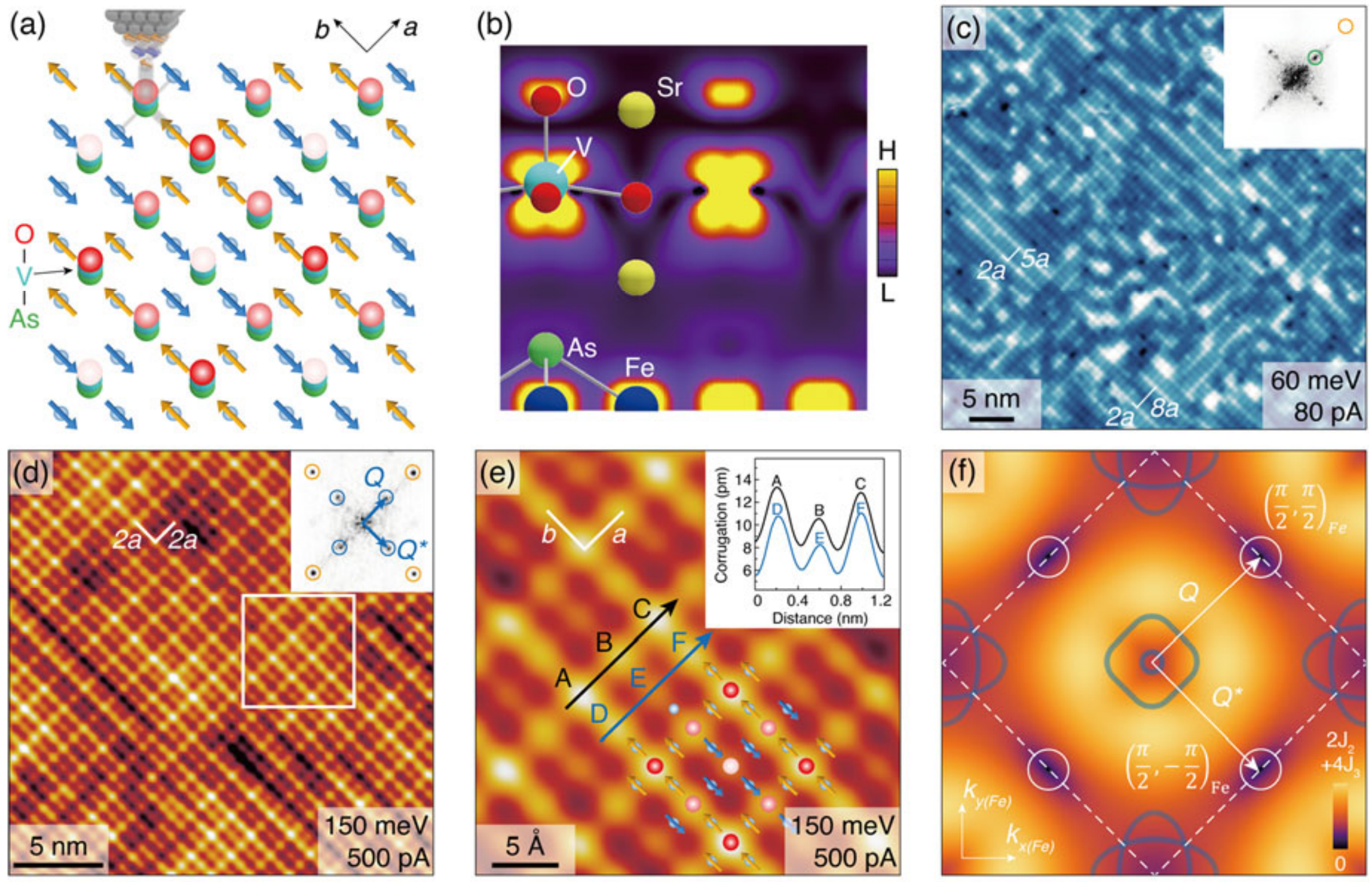

FIG. 2. (a) The structure of the Fe magnetic moments in the $C_4$ (plaquette) order. Each red dot represents the oxygen at the top of each vertical O-V-As atomic chain acting as the tunneling path. (b) The theoretical electron density plot near the Fermi level (integrated over $[-50,0]$ meV). (c) A typical 4.6 K topograph taken with W tip showing quasi-$C_2$ SRs with random orientations. (d) A spin-polarized STM image taken at low junction resistance with nearly-in-plane polarized Cr tip showing the induced $C_4$-symmetric (2×2) order. The orange, green, and blue circles in the Fourier-transformed images (the insets) of (c) and (d) indicate $|\boldsymbol{q}| = 2\pi/a$ (Bragg peaks), $3\pi/4a$ and $\pi/a$ respectively. (e) The magnified view of the area in a white square in (d), with the (2×2) magnetic unit cells with a $C_4$ plaquette spin model overlayed. Its inset shows the cross-sections along the black and blue arrows. (f) The spin wave dispersion of the $C_4$ plaquette order and its two momentum transfer vectors ($\boldsymbol{Q}$ and $\boldsymbol{Q^*}$) from localized moment picture [8], shown together with the ARPES-based Fermi surfaces (dark curves) [33].

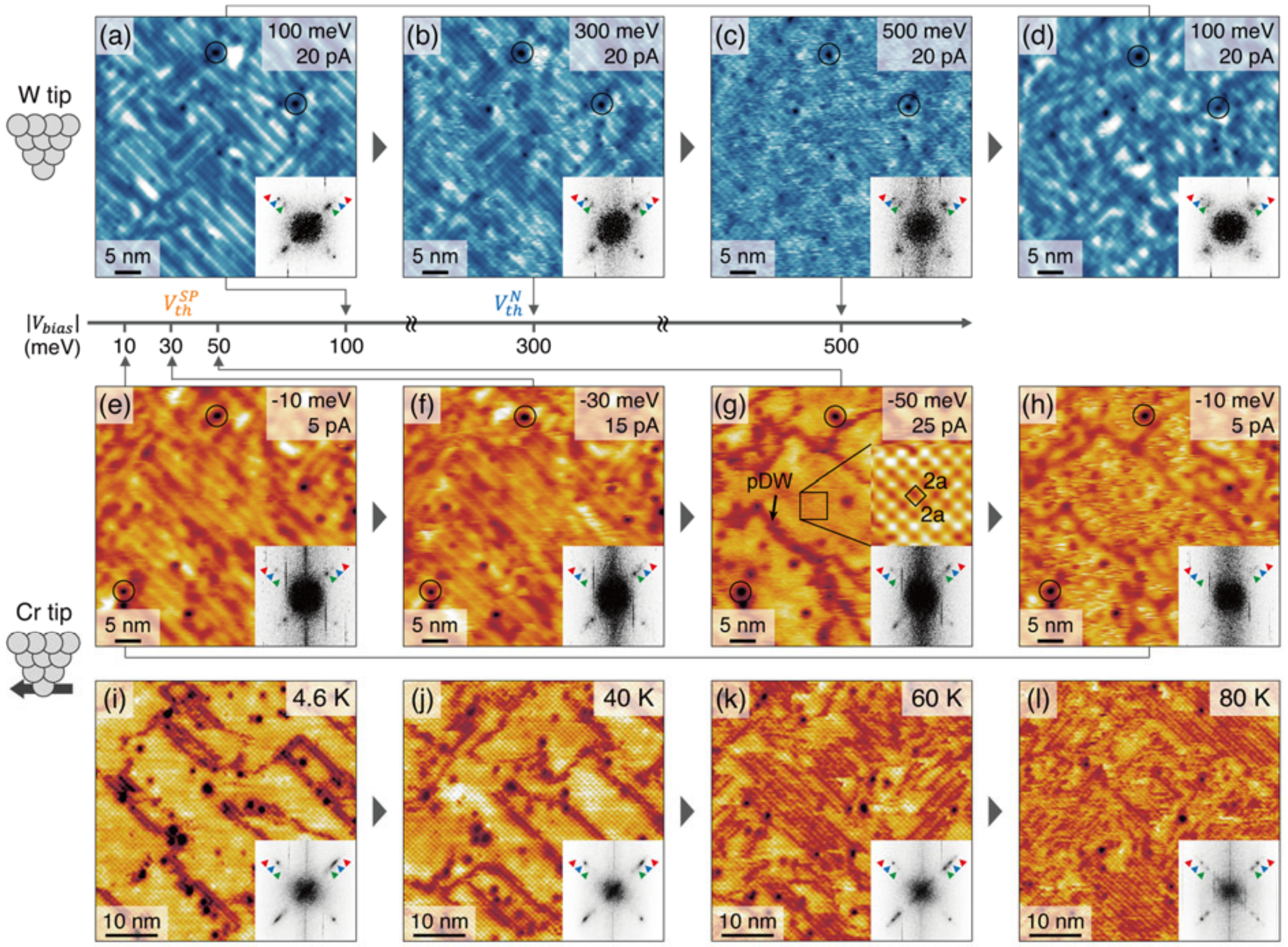

FIG. 3. (a)-(d) Bias dependence of W-tip topograph images showing a threshold voltage for SR fluctuation near $V_{th}^{N} \approx$ 300 meV. (e-h) Bias dependence of Cr-tip topograph images. $C_4$ symmetric (2×2) domains and phase domain walls (pDW) (g) are induced at a significantly lower threshold ($V_{th}^{SP} \approx$ 30 meV) indicating a qualitatively different final state from that obtained with the W tip (d). Inset in (g) is taken at slightly higher junction conductance [-50 mV, 100pA]. In all the FFT insets, the blue (red, green) arrows correspond to $|\boldsymbol{q}| = \pi/a_0$ ($5\pi/4a_0$, $3\pi/4a_0$). (i)-(l) Temperature-dependent Cr-tip topographs taken at bias [-50 mV, 100pA] showing erasure of the $C_4$ order beyond the Fe magnetic ordering temperature.

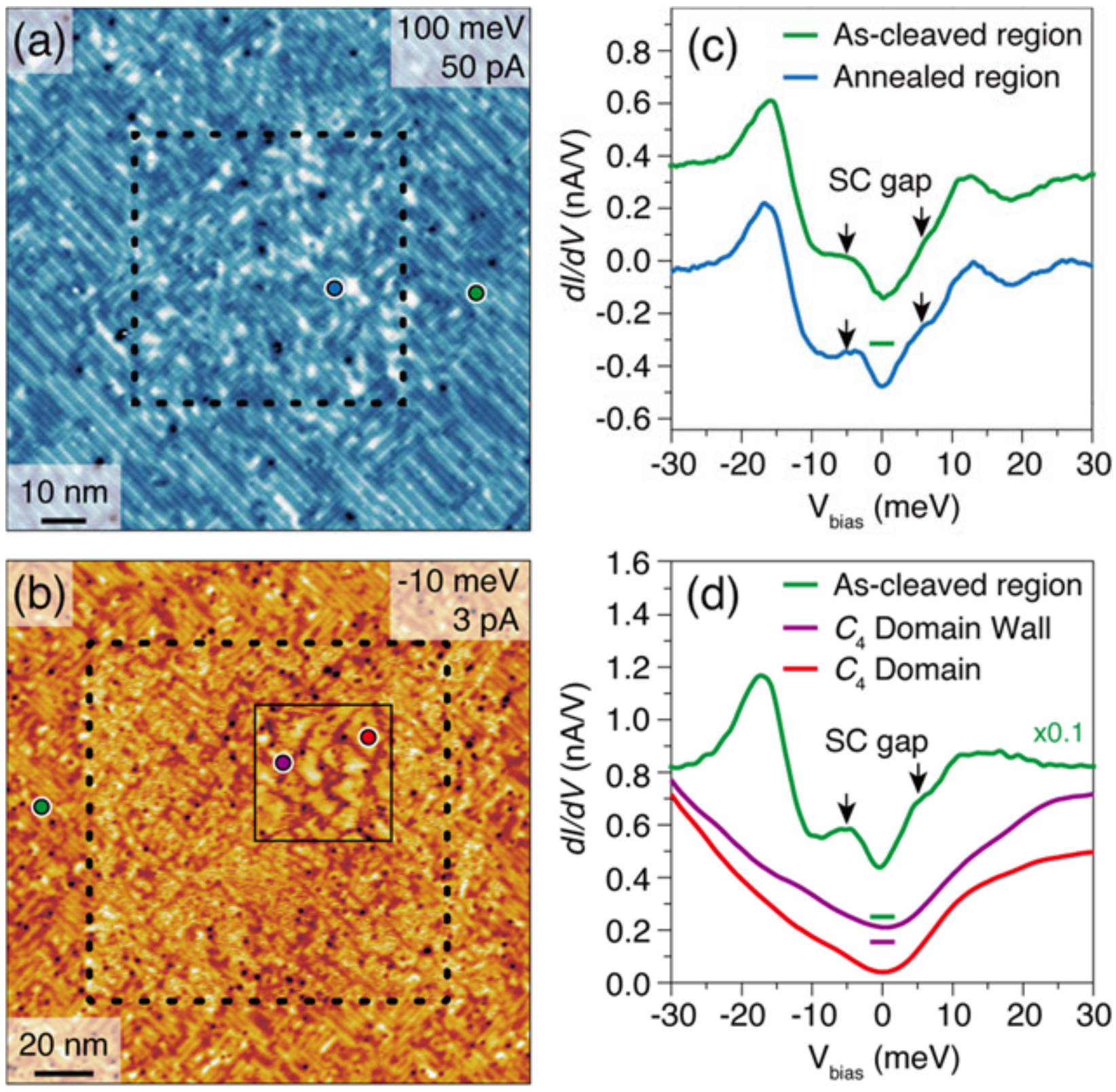

FIG. 4. (a)[(b)] W-tip (Cr-tip) topograph with bias conditions below threshold $V_{th}^{N}$ ($V_{th}^{SP}$) taken after the higher bias scans shown in Fig. 3(a)-(c) [Fig. 3(e)-(g)] performed only in the area inside the dotted square. The inset image in the solid square in (b) is a Cr-tip topograph with higher bias above $V_{th}^{SP}$ showing the domains and the domain walls more clearly (See Fig. S8). (c)[(d)] shows the tunneling spectra measured at marked positions with corresponding marker colors in (a) with a W tip [(b) with a Cr tip]. The W tip spectra was measured at bias [40 meV, 300 pA]. The Cr tip spectra was measured at set point of [50 meV, 120 pA] inside the $C_4$ region, and at set point of [30 meV, 5 pA] in the pristine region with larger averaging time to avoid inducing $C_4$ state during *dI/dV* measurement. All the data are taken at 4.6 K.

# Switching Magnetism and Superconductivity with Spin-Polarized Current in Iron-Based Superconductor

Seokhwan Choi,[1] Hyoung Joon Choi,[2] Jong Mok Ok,[3,4] Yeonghoon Lee,[1] Won-Jun Jang,[1,5,‡] Alex Taekyung Lee,[6] Young Kuk,[7] SungBin Lee,[1] Andreas J. Heinrich,[8,9] Sang-Wook Cheong,[10] Yunkyu Bang,[11] Steven Johnston,[12] Jun Sung Kim,[3,4] and Jhinhwan Lee[1]*

*E-mail: jhinhwan@kaist.ac.kr

## 1. Theoretical study on metastability of $C_4$ magnetism induced by spin-polarized current

We used a first-principles pseudopotential method with the generalized gradient approximation to the density functional theory (DFT), as implemented in the SIESTA code. We used pseudoatomic orbitals to expand the electronic wavefunctions and semicore norm-conserving pseudopotentials for Sr, V, and Fe, as implemented in the SIESTA code. We used experimental crystal structure of $Sr_2VO_3FeAs$, which was measured by x-ray diffraction as shown in Table 1 of Ref. 22. We included Sr 4s, 4p, 5s, V 3s, 3p, 3d, 4s, Fe 3s, 3p, 3d, 4s electrons as valence electrons. To describe the single-stripe and plaquette antiferrromagnetic orderings, we used a 2×2×1 supercell containing eight Fe atoms in a single FeAs layer.

We obtained the total energy of the system for the single-stripe and plaquette antiferromagnetic orderings of Fe magnetic moments. As shown in Fig. S1, the single-stripe antiferromagnetic ordering in the Fe layer is more stable than the plaquette ordering by about 80 meV/Fe when the sample has an equal number of spin-up and spin-down electrons, that is, the net ferromagnetic moment is zero. For V layers, we obtained that ferromagnetic ordering is stable within each V layer, with negligible magnetic coupling between adjacent V layers.

Then, in order to examine the experimental situation of having spin-polarized current injected from the STM tip to the sample, we considered breaking the number balance of the spin-up and spin-down electrons in our DFT calculations. As shown in Fig. S1, when we broke the number balance of the spin-up and spin-down electrons, that is, the net ferromagnetic moment was introduced to the system, the energy difference between the single-stripe and plaquette orderings decreased gradually. Finally, the plaquette ordering becomes more stable than the single-stripe orderings when the net ferromagnetic moment is larger than about one Bohr magneton per formula unit. This result shows that the imbalance of the spin-up and spin-down electrons may stabilize the plaquette ordering and it supports our STM results with spin-polarized current, although it has only qualitative validity in the sense that such a large net moment might not occur in our STM experiment and, furthermore, the experimental situation is more complicated, with the presence of Cr atoms of the STM tip, strong local electric field from the tip, and deviation from equilibrium with net current.

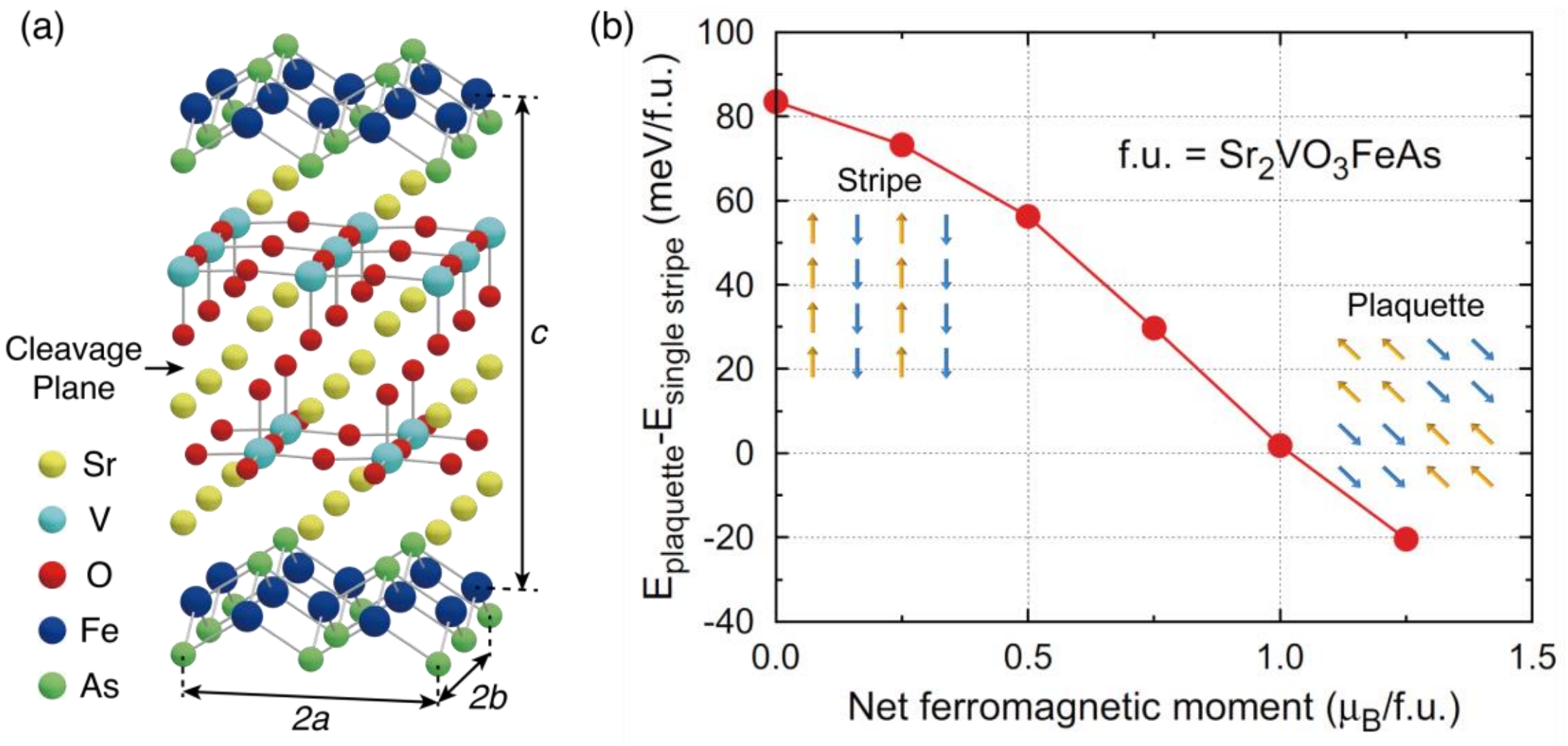

FIG. S1. (a) A structural model of tetragonal $Sr_2VO_3FeAs$ lattice showing the calculation unit cell. (b) The total-energy difference between the plaquette and single-stripe antiferromagnetic orderings as a function of net ferromagnetic moment. The energy difference is in meV per formula unit ($Sr_2VO_3FeAs$) and the net ferromagnetic moment is in Bohr magneton per formula unit.

## 2. Roles of V and Fe in the magnetism and the superconductivity of $Sr_2VO_3FeAs$

The main controversy and debating issue about the $Sr_2VO_3FeAs$ compound is the role of $VO_2$ layers and V moment. Early on, there were several works (Refs. 21 and 22) claiming for V magnetic moment ordering and its possible role for superconducting pairing. However, recent nuclear magnetic resonance (NMR) work (Ref. 29) has measured $^{51}$V-NMR and $^{75}$As-NMR signals over the wide temperature range and revealed following information: (1) $^{51}$V-NMR signal (Knight shift) remains completely unchanged from 200 K down to a few K (far below $T_c$ ~33 K). (2) On the contrary, $^{75}$As-NMR signals (both Knight shift and $1/T_1T$) detect all important phase transitions such as the Fe magnetic ordering temperature (~ 50 K) and the superconducting transition (~ 33 K). This experimental results lead to the conclusion that the $VO_2$ layers and V moments don't play any active roles either for magnetism or for superconductivity except for inducing the frustration by coupling with the FeAs layer.

Indeed, based on the DFT calculations, the V *d*-electrons provide a large spectral density in the range of [−50,0] meV and they participate to the Fermi surfaces together with the Fe *d*-electrons. However, Ref. 21 has shown with his DFT calculations that the electronic spectra near Fermi level, provided by the $VO_2$ layers and the FeAs layers, primarily work separately and the coupling between them is a secondary correction. For example, superconductivity primarily occurs on the FeAs layers as the portion of the Fermi surface made of Fe *d*-electrons determines the superconducting pairing and the portion made of V *d*-electrons passively follows. This theoretical picture is in accord with the conclusion of this paper.

## 3. Preparation of Cr tip for SPSTM

The spin-polarized Cr cluster tip was prepared by collecting Cr atoms on the apex of the W tip by controlled field emission with parameters depending on the sharpness of the base W tip. The Cr cluster tip was then tested for proper in-plane spin-polarization by observing multiple levels of differential conductance at set point bias near −50 meV on multiple antiferromagnetic terraces with identical orientation on a stepped Cr surface as shown in Fig. S2.

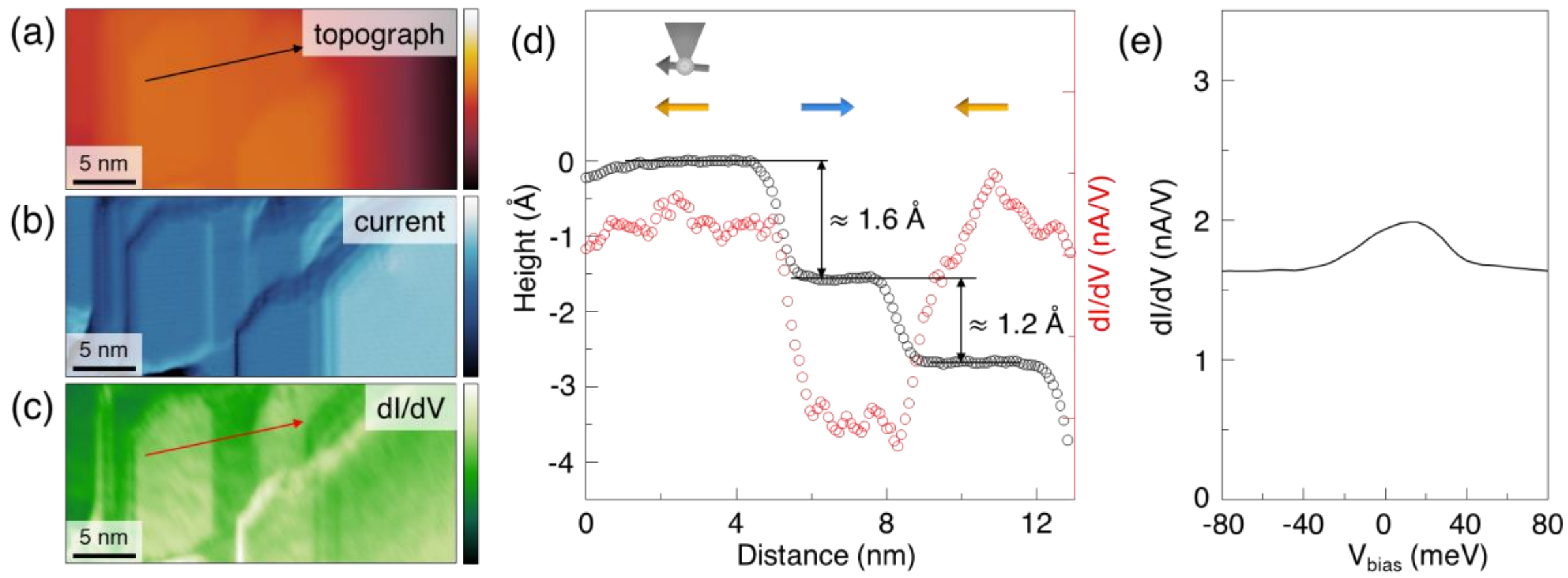

FIG. S2. A sputter-annealed Cr(001) surface topograph (a), and simultaneously taken current image (b), and *dI/dV* image (c), at bias condition of (−50 mV, 200 pA). (c) The spin-contrast shown in cross-sections of topograph and *dI/dV* along a line marked with arrows in (a) and (c). (e) *dI/dV* spectrum taken with bias setpoint 1 nA, −200 meV on Cr(001) surface measured by Cr tip.

## 4. SPSTM characteristics of Plaquette and other possible orders in $Sr_2VO_3FeAs$

In the Cr-cluster tip SPSTM imaging, the vertical tunneling conduction channel made of O-V-As to the Fe layer equally samples the spin polarization of the four Fe atoms neighboring of each As atom. When every four Fe spins underneath a particular O-V-As chain are parallel (antiparallel) to the Cr tip's spin polarization, the top layer O atom will look bright (dark) in the SPSTM topograph. On the other hand, if the four Fe spins underneath an O-V-As chain are grouped into two roughly parallel spins and two roughly antiparallel spins, they will possess neutral brightness just at the average of the brightness of the bright and the dark O atoms.

In this condition, the plaquette (ODS) order in Fe layer generates a (2×2) magnetic unit cell with characteristic intra unit cell pattern as simulated and shown in Fig. S3 (a). In contrast, the DDS (diagonal double stripe) order, the Néel order, the single stripe order, and the two kinds of spiral orders that may appear in the spin-order phase diagram for Heisenberg model have qualitatively different magnetic unit cells as shown in Fig. S3 (b)-(f).

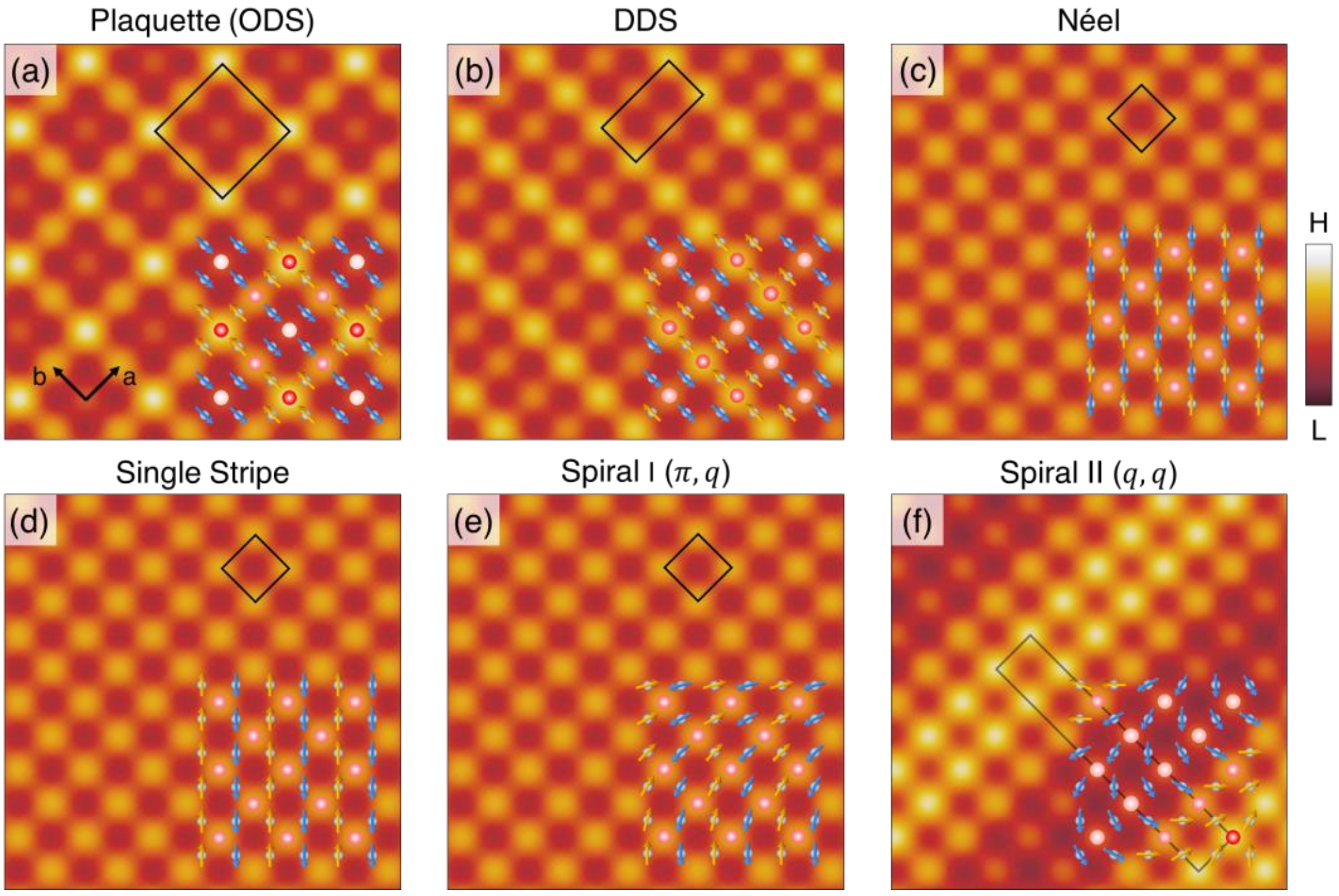

FIG. S3. Simulated SPSTM topographs for various possible magnetic orders in the tunneling geometry of $Sr_2VO_3FeAs$ in case of no surface reconstruction. (a) Plaquette (orthogonal double stripe, ODS) order. (b) DDS (diagonal double stripe) order. (c) Néel order. (d) Single stripe order. (e) Spiral order I with wavevector $(\pi, \boldsymbol{q})$. (f) Spiral order II with wavevector $(\boldsymbol{q}, \boldsymbol{q})$.

## 5. Simulation of phase domain walls of $C_4$ magnetic order induced by scanned spin-polarized current

Using the Landau-Lifshitz-Gilbert (LLG) equation and the tunneling geometry of $Sr_2VO_3FeAs$, we have simulated the SPSTM image showing the phase domains and domain walls of the Plaquette magnetic order in the Fe layer, assuming the tunneling geometry of $Sr_2VO_3FeAs$ and no surface reconstruction [38]. The similarity with the experiment is an indirect evidence of the induced $C_4$ state being the $C_4$ plaquette order in the Fe layer.

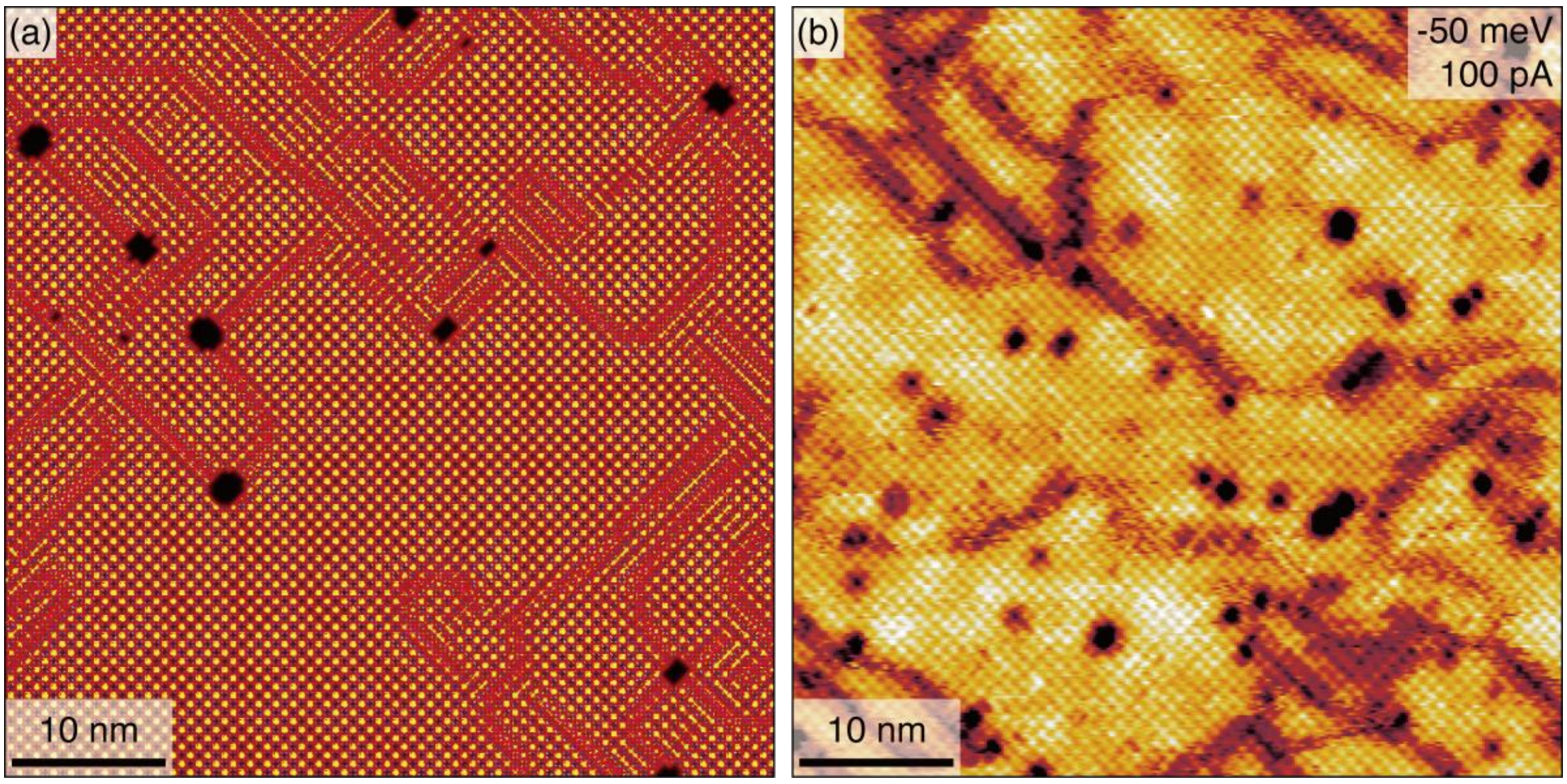

FIG. S4. (a) The LLG equation-based simulation result assuming the tunneling geometry of $Sr_2VO_3FeAs$ and no surface reconstruction. (b) The experimental Cr-tip image of the same sized area showing dark phase domain walls.

## 6. $C_4$ (2×2) phase domain walls visualized by spatial lock-in technique

The domain walls of the plaquette (ODS) order can be determined by first detecting the (2×2) spatial modulation phases ($\phi_a(\vec{r})$,$\phi_b(\vec{r})$) with a technique similar to the time-domain phase detection method known as the lock-in technique. To detect the modulation phase $\phi_a(\vec{r})$ in $a$-direction, we choose the Bragg peak position $\boldsymbol{q}_{(\pi,0)}$ corresponding to $(\pi, 0)$ in the Fourier Transform (FT) of the SPSTM topograph. We then generate two arrays *S* and *C* with identical size as the original topograph *T*, where *S* and *C* are filled with $\sin(\boldsymbol{q}_{(\pi,0)} \cdot \boldsymbol{r})$ and $\cos(\boldsymbol{q}_{(\pi,0)} \cdot \boldsymbol{r})$ respectively. Pixel-by-pixel multiplication of *T* and *C* denoted as $C_T$ (and *T* and *S* denoted as $S_T$) contains fast spatial modulations with modulation wave vector near $2\boldsymbol{q}_{(\pi,0)}$ and slow spatial modulations with wave vector near 0. As with the time-domain lock-in technique, we filter out the fast $2\boldsymbol{q}_{(\pi,0)}$ modulations with a spatial low pass filter with cut off wave vector $\sim\boldsymbol{q}_{(\pi,0)}$ and denote them $\langle C_\mathrm{T}\rangle$ and $\langle S_\mathrm{T}\rangle$. Such $\langle C_\mathrm{T}\rangle$ and $\langle S_\mathrm{T}\rangle$ contain information of $A\cos\phi_a(\vec{r})$ and $A\sin\phi_a(\vec{r})$ respectively and the phase $\phi_a(\vec{r}) = \tan^{-1}(\langle C_\mathrm{T}\rangle, \langle S_\mathrm{T}\rangle)$ can be defined at every pixel of the topograph as shown in Fig. S5 (c). The domain walls (red curves in Fig. S5 (b)) determined by the spatial modulation phase shift in $a$-direction can then be defined as the collection of pixels with abrupt reversal (change by ~$\pi$) of phase $\phi_a(\vec{r})$ within a magnetic unit cell distance from the pixel. Applying the identical method starting with the Bragg peak position $\boldsymbol{q}_{(0,\pi)}$ will generate $\phi_b(\vec{r})$ (Fig. S5 (d)) and the blue domain walls (blue curves in Fig. S5 (b)). The purple domain wall results from overlapping red and blue domain walls.

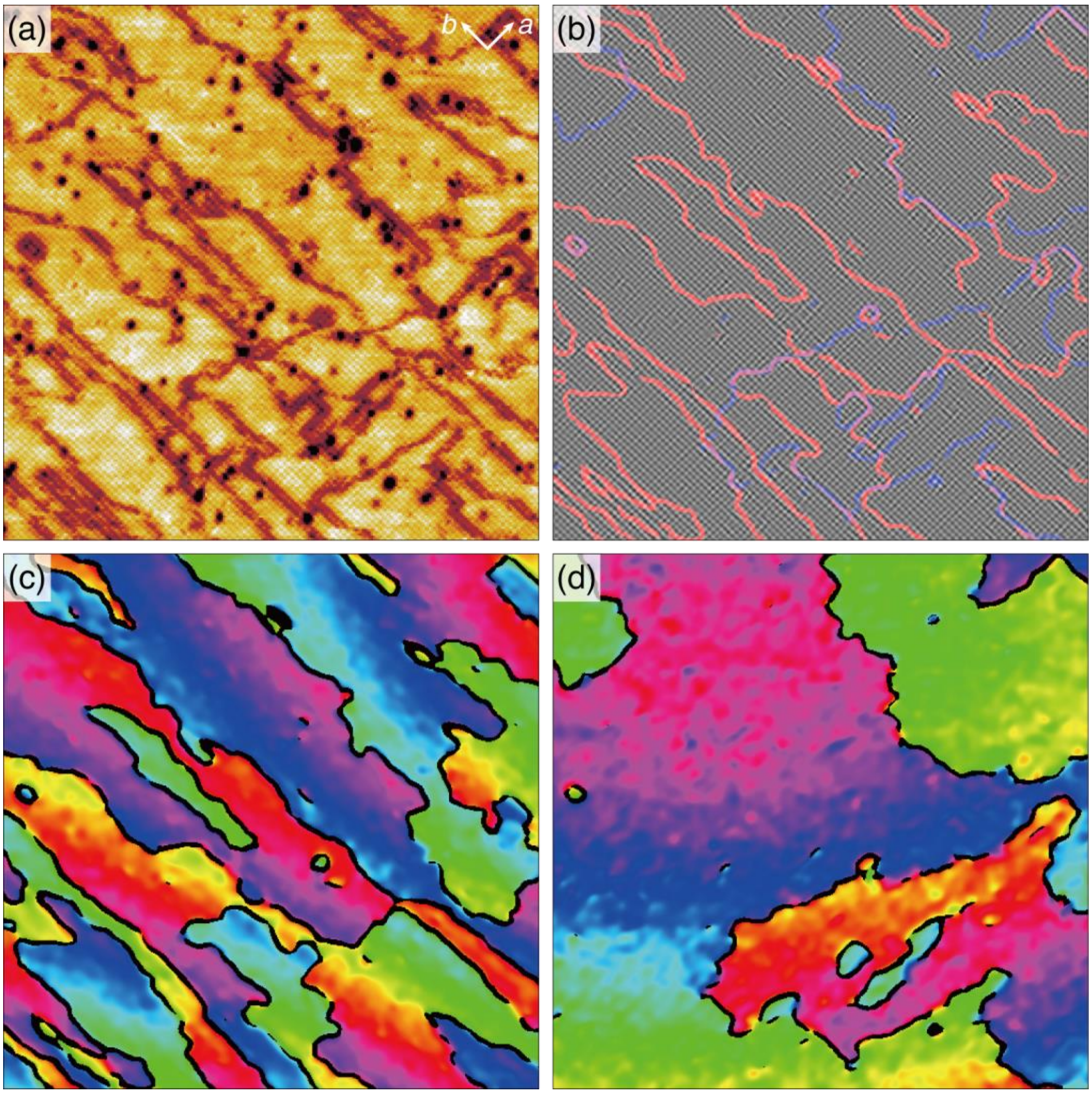

FIG. S5. (a) Large area (71 nm x 71 nm) topograph taken at 4.6 K with spin-polarized Cr tip at bias condition of (−50 meV, 100 pA) on $Sr_2VO_3FeAs$. (b) Image of the automatically detected domain walls showing the (2×2) atomic modulations in each domain. The bright dots in each domain correspond to the brightest atom in the (2×2) unit cell. (c,d) The spatial modulation phase maps of $\phi_a(\vec{r})$ (c) and $\phi_b(\vec{r})$ (d), used to automatically detect the two types of domain walls. The piezo-creep-induced lattice distortion results in the slow variations of the phases over the field of view, which do not affect the domain wall detection algorithm relying on the abrupt phase change by $\pi$ within the width ($\sim 2a_0$) of the domain walls.

## 7. Near degeneracy of metastable $C_4$ order driven by spin-polarized tunneling current

In case of near degeneracy in metastable $C_4$ order made temporarily stable by spin-polarized tunneling current, we can expect to observe telegraphic noise in the tunneling current whose fluctuation rate decreases with decreasing bias voltage until it’s too small compared with the energy difference between the two metastable states. As shown in Fig. S6 below, once the plaquette order is driven by strong spin-polarized tunneling current with bias voltage well over $V_{th}^{SP} \approx$ 30 meV, significant amount of telegraphic noise begins to be observed which becomes clearer as the bias voltage is brought down below 30 meV due to the fluctuation rate decreasing below the preamp response frequency. The hysteresis of the topography as well as the telegraphic noise fluctuation rate due to metastability of the $C_4$ order is clearly visible as a function of the bias voltage.

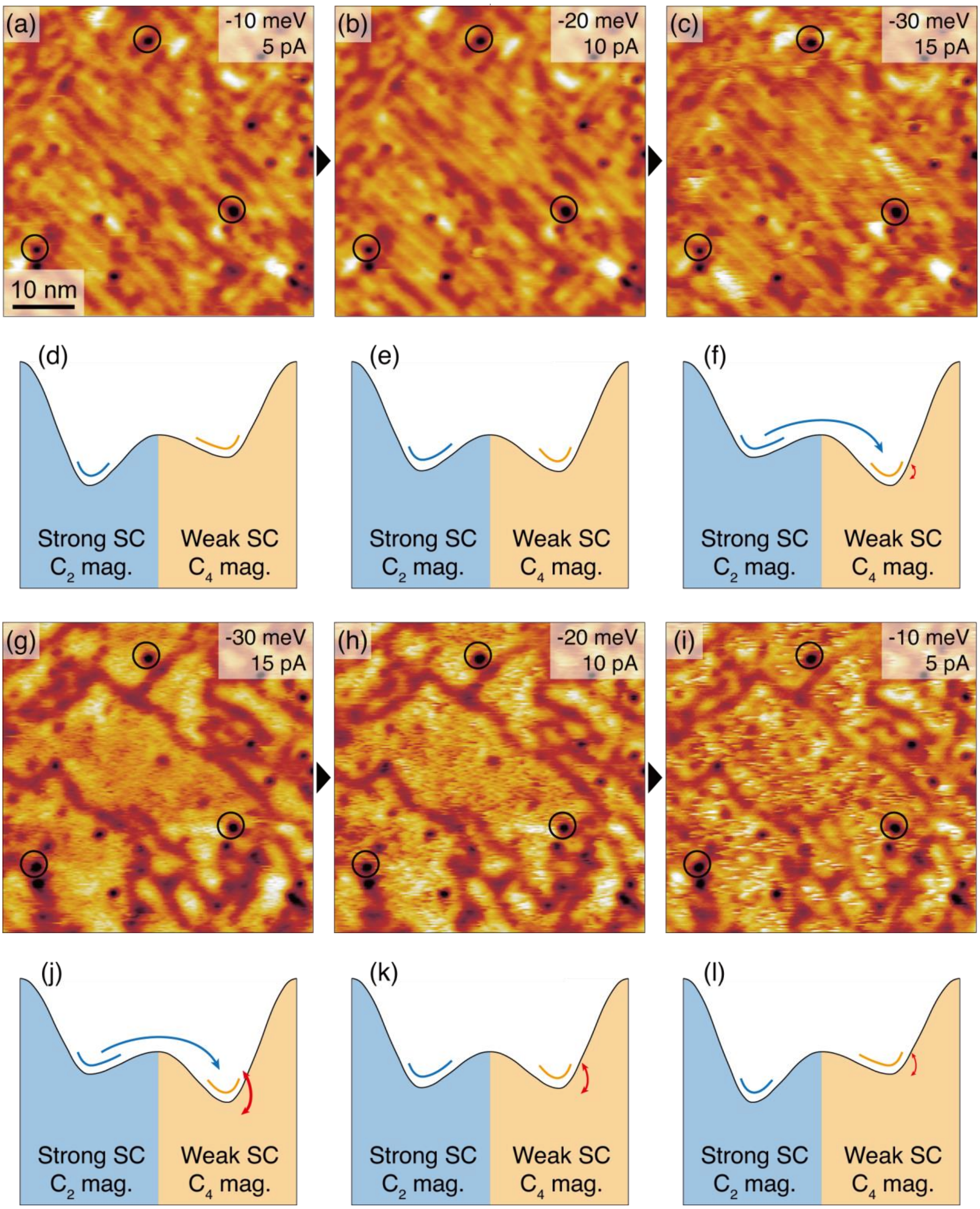

FIG. S6. Evidence of near degeneracy of the metastable $C_4$ order driven by spin-polarized tunneling current. Once the bias voltage of the strong spin-polarized tunneling current is driven beyond $V_{th}^{SP} \approx$ 30 meV, the $C_4$ order appears near the tip and is retained even when the bias voltage is reduced well below $V_{th}^{SP}$ again. With the appearance of $C_4$ order, telegraphic noise also appears and its rate increases with bias voltage and current which is an evidence of near degeneracy of the $C_4$ order.

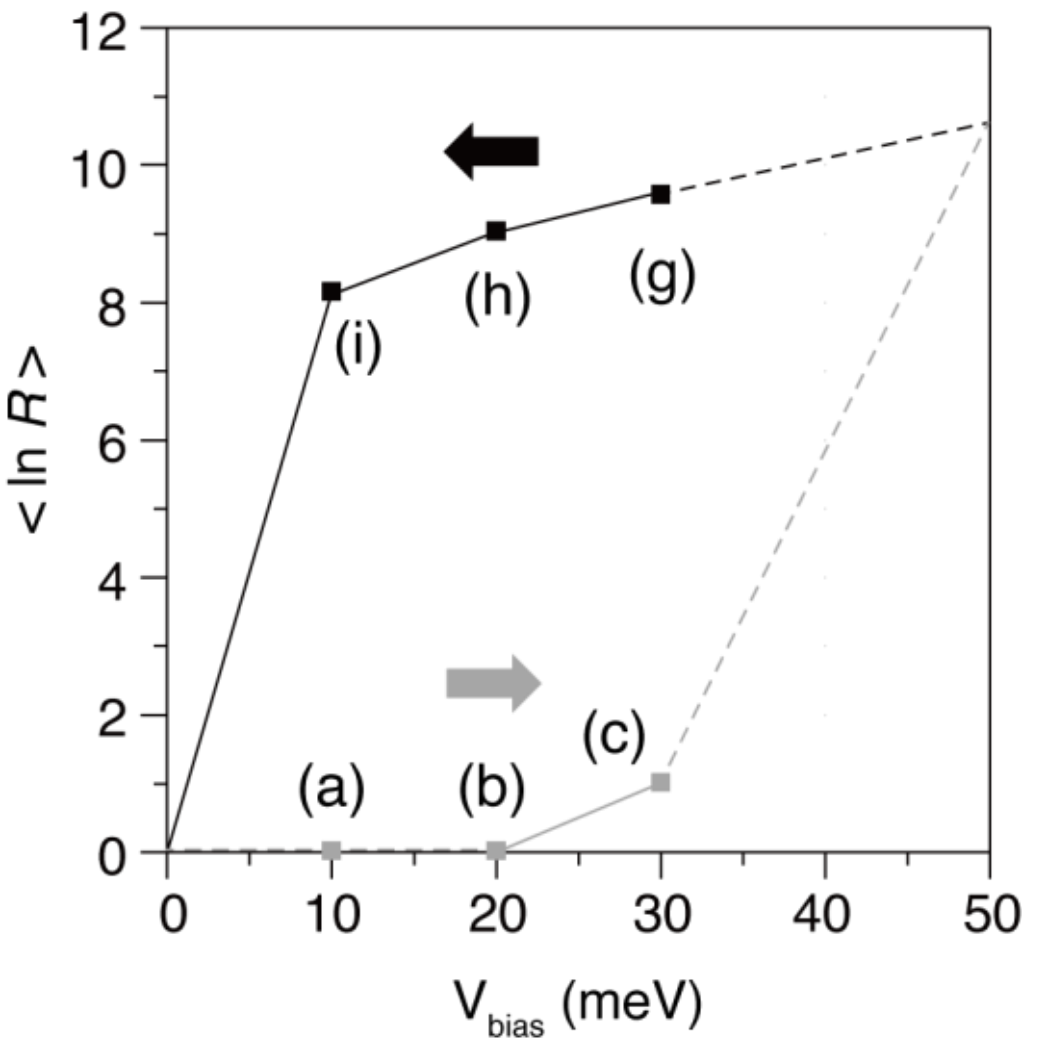

FIG. S7. The rate of telegraphic noises showing bias hysteresis due to the tiny potential barriers of several meV's between the nearly degenerate states of the metastable $C_4$ order driven by spin-polarized tunneling current. (a)-(i) indicate the corresponding panels in Fig. S6.

The switching mechanism shown in Fig. S1 connects the tunneling current level, rather than the bias voltage level, to the change in the plaquette order energy relative to the energy of the $C_2$ magnetic state. However, due to the existence of the energy barrier, there should be bias voltage threshold too, and both the tunneling current level and the bias voltage level will have threshold levels slightly depending on each other. As shown in Fig. S9, we acquired a set of additional Cr tip topograph images with increasing bias voltage level when the tunneling current level is fixed at 20 pA. We can compare it with the extended version of Fig. 3 shown in Fig. S8 below. From the fact that the flat bright areas made of $C_4$ magnetic order are increasing in the order of [-30 mV, 15 pA] (Fig. S8(c), green) < [-30 mV, 20 pA] (Fig. S9(e), blue) < [-40 mV, 20 pA] (Fig. S8(d), red), we can conclude that both the tunneling current level and the bias voltage level play significant positive roles in the probability of switching into the $C_4$ state.

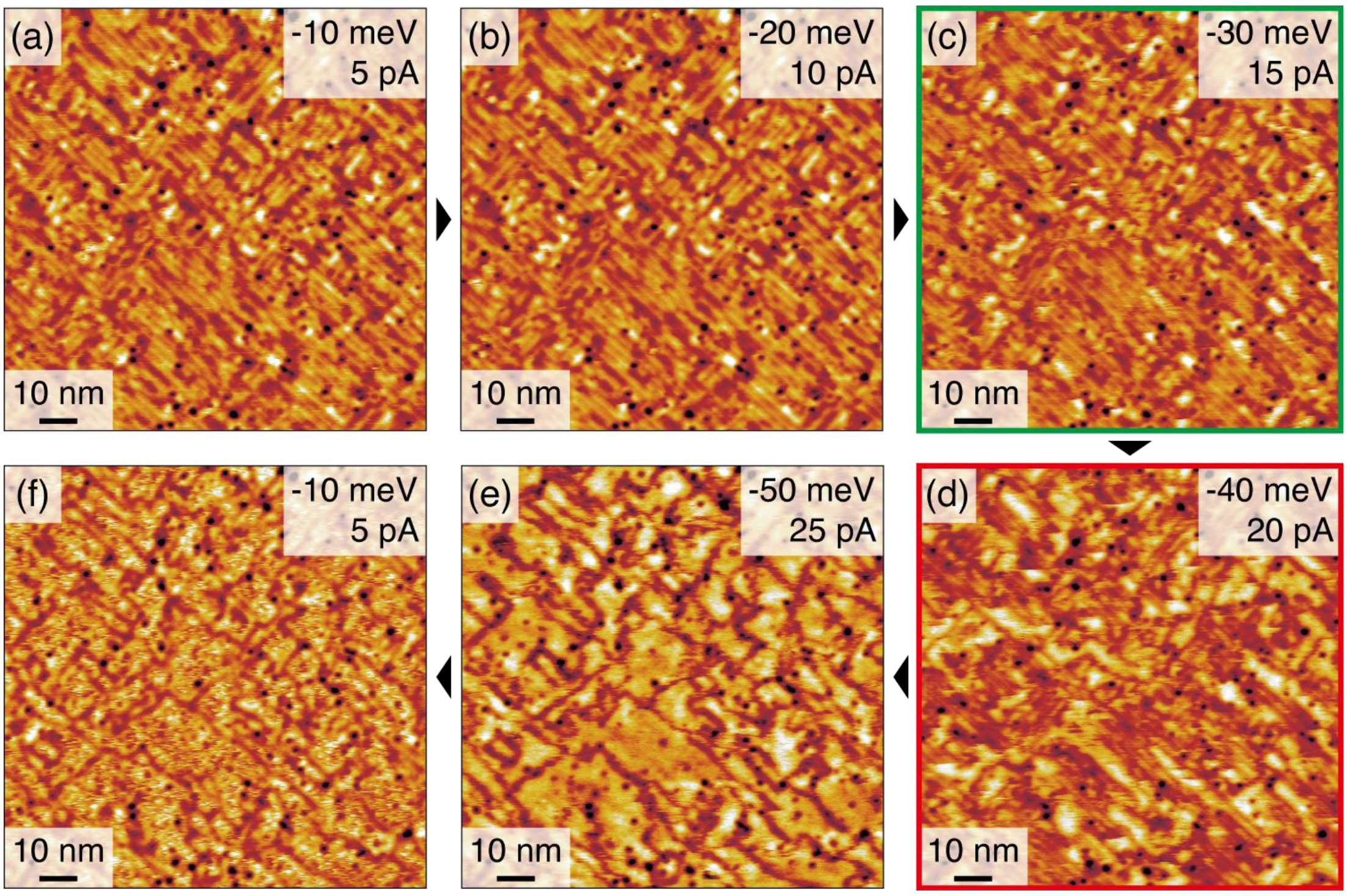

FIG. S8. Cr tip SPSTM topographs measured as the bias voltage is varied proportionally with the tunneling current level.

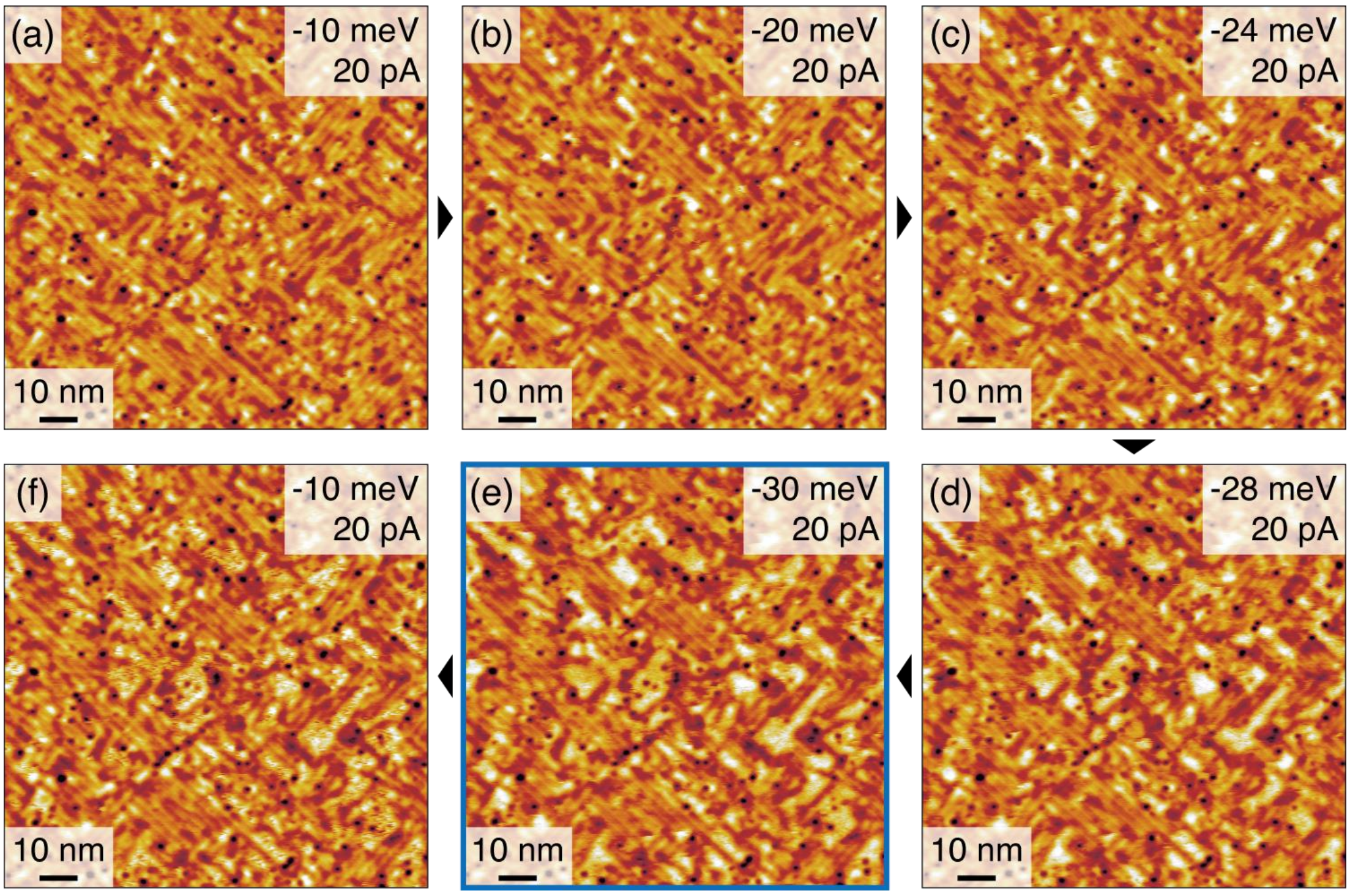

FIG. S9. Cr tip SPSTM topographs measured as the bias voltage is varied at a fixed tunneling current level of 20 pA.

We have performed the thermal cycle experiment showing the erasure of the $C_4$ magnetic order beyond 60 K and put the related figure as shown in Fig. S10. The threshold temperature of 60 K agrees well with the Fe magnetic ordering temperature observed in NMR experiment in Ref. 29.

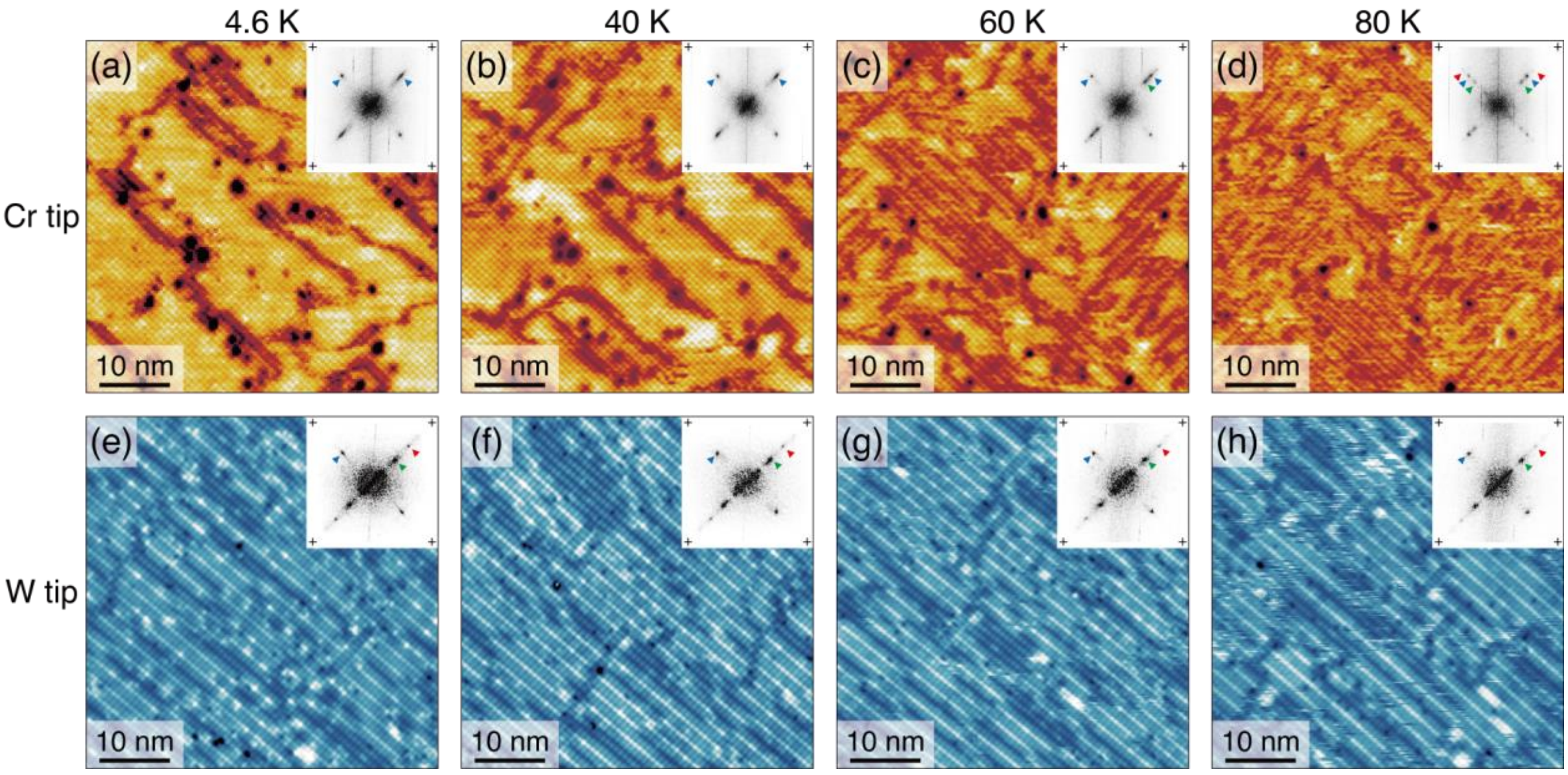

FIG. S10. Erasure of $C_4$ magnetic order by thermal cycle. The Cr tip SPSTM images (a)-(d) are taken at the respective temperatures marked above each panel in the order of increasing temperature. The W tip topographs show no qualitative change. The setpoint conditions are [-50 meV, 100 pA] for Cr tip images and [100 meV, 300 pA] for W tip images. In the FFT insets, the blue (red, green) arrows correspond to $|\boldsymbol{q}| = \pi/a_0$ $(5\pi/4a_0,\ 3\pi/4a_0)$.

We also performed magnetic field dependent SPSTM imaging of the $C_4$ magnetic order using a 7 T magnet in our STM. Due to the antiferromagnetic nature of the $C_4$ magnetic order, there was virtually no change in the $C_4$ magnetic order up to 7 T as shown in Fig. S11 below. Considering the 60 K temperature threshold of the above thermal cycle experiment, we may need several times larger magnetic field to achieve qualitative change in the $C_4$ state.

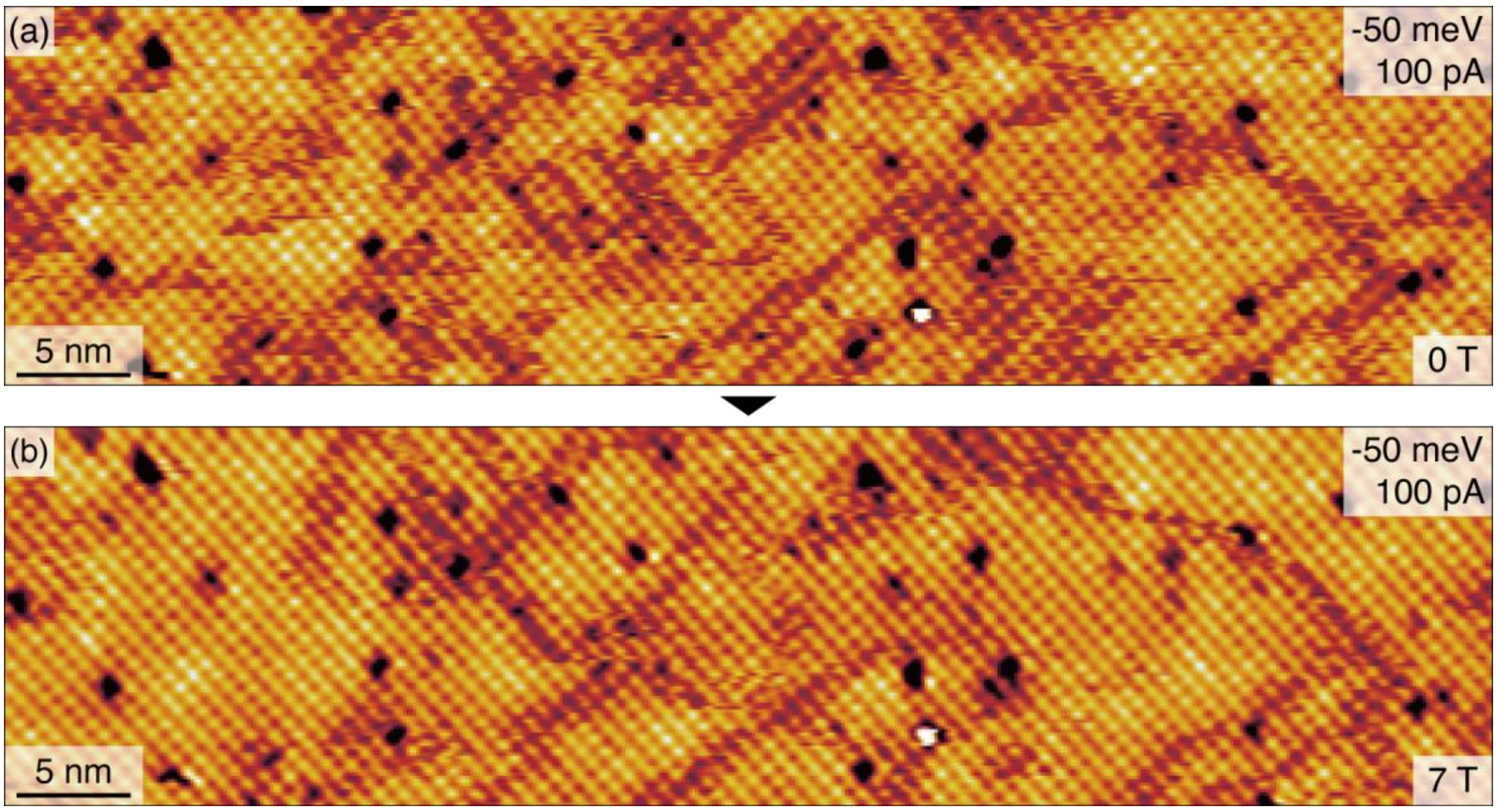

FIG. S11. Magnetic field dependence of Cr tip SPSTM topographs. There is little qualitative change up to 7 T, except for the tunneling-current-induced phase domain wall motions that occur irrespective of the magnetic field [38].

## 8. Conceptual design of device switching magnetism and superconductivity using spin-polarized and unpolarized current.

The principle of switching magnetism and superconductivity by spin-polarized and unpolarized currents as demonstrated by our research may be used to build a circuit device made of a single $Sr_2VO_3FeAs$ layer sandwiched by a ferromagnetic electrode and a normal metal electrode as shown in Fig. S12 below.

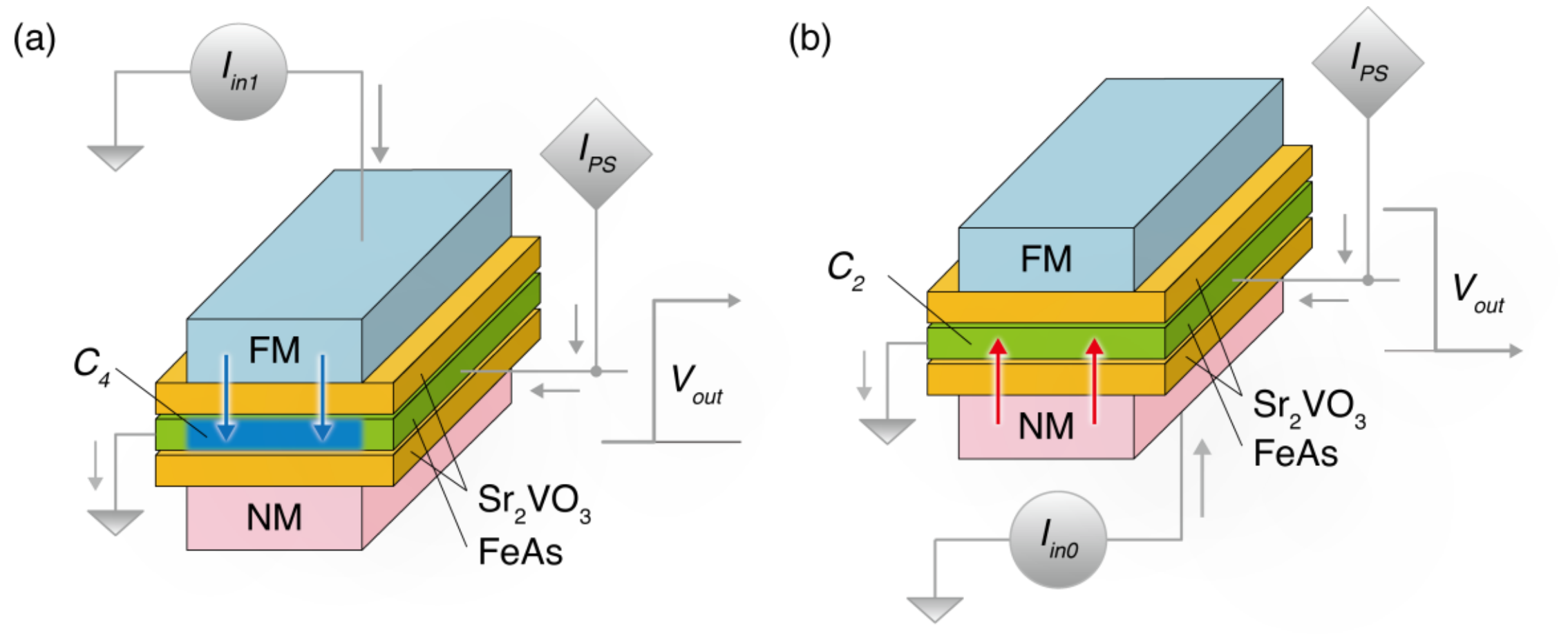

FIG. S12. Conceptual diagrams of electronic device switching magnetism and superconductivity in the channel using currents from (a) ferromagnetic (FM) and (b) normal metal (NM) electrodes.